\documentclass[twocolumn,showpacs,preprintnumbers,amsmath,amssymb]{revtex4}

\usepackage{graphicx}
\usepackage{dcolumn}
\usepackage{bm}
\usepackage{amsmath}

\begin{document}
\title{Phase Synchronization of non-Abelian Oscillators on Small-World Networks}
\author{Zhi-Ming Gu$^1$}
\author{Ming Zhao$^2$}
\author{Tao Zhou$^2$}
\author{Chen-Ping Zhu$^1$}
\author{Bing-Hong Wang$^2$}
\affiliation{%
$^1$College of Science, Nanjing University of Aeronautics and Astronautics, Nanjing 210016, PR China\\
$^2$Department of Modern Physics, University of Science and
Technology of China, Hefei 230026, PR China}%

\date{\today}

\begin{abstract}
In this paper, by extending the concept of Kuramoto oscillator to
the left-invariant flow on general Lie group, we investigate the
generalized phase synchronization on networks. The analyses and
simulations of some typical dynamical systems on Watts-Strogatz
networks are given, including the $n$-dimensional torus, the
identity component of 3-dimensional general linear group, the
special unitary group, and the special orthogonal group. In all
cases, the greater disorder of networks will predict better
synchronizability, and the small-world effect ensures the global
synchronization for sufficiently large coupling strength. The
collective synchronized behaviors of many dynamical systems, such as
the integrable systems, the two-state quantum systems and the top
systems, can be described by the present phase synchronization
frame. In addition, it is intuitive that the low-dimensional systems
are more easily to synchronize, however, to our surprise, we found
that the high-dimensional systems display obviously synchronized
behaviors in regular networks, while these phenomena can not be
observed in low-dimensional systems.
\end{abstract}

\pacs{89.75.Hc, 05.45.Xt}

\maketitle

\section{Introduction}
Synchronization is observed in many natural, social, physical and
biological systems, and has found applications in a variety of
fields \cite{Strogatz2003}. The large number of networks of
coupled dynamical systems that exhibit synchronized states are
subjects of great interest. As a theoretical paradigm, the
Kuramoto model is usually used to investigate the phase
synchronization among nonidentical oscillators
\cite{Kuramoto1984,Pikovsky2001,Acebron2005}. However, some real
physical systems that display synchronized phenomenon (e.g. the
coupled double planar pendulums, the two-state quantum system,
etc.) can not be properly described by the simply Kuramoto model.
Actually, the Kuramoto oscillator is equal to the left-invariant
flow on the simplest nontrivial Lie group $U(1)=\{e^{i\zeta}|\zeta
\in R\}$ (see the Refs. \cite{Schutz1980,Brocker1985} for the
fundamental knowledge of Lie group). Accordingly, in this paper,
by extending the concept of Kuramoto oscillator to left-invariant
flow on general Lie group, we investigate the generalized phase
synchronization on networks. In particular, the left-invariant
flows corresponding to non-Abelian Lie groups are named
non-Abelian oscillators, which can be applied on some non-Abelian
dynamical systems like quantum systems.

This paper is organized as follow: In section 2, the concept of
synchronization of non-Abelian oscillators is introduced. In section
3, the analyses and simulations on some typical examples are given,
including the $n$-dimensional torus $T^n$, the identity component of
3-dimensional general linear group $GL(3,R)$ (i.e. $GL_0(3,R)$), the
special unitary group $SU(2)$ and the special orthogonal group
$SO(3)$. Finally, we sum up this paper and discuss the relevance of
phase synchronization of non-Abelian oscillators to the real world
in section 4. Some mathematic remarks are given in the Appendix.

\section{Synchronization of non-Abelian oscillators}

First of all, we review the simple oscillator:
\begin{equation}
e^{i\left( {\omega t + \theta } \right)} = Ce^{i\omega t}, \quad C
= e^{i\theta }.
\end{equation}
It is obvious that $x(t) = \omega t + \theta $ is the solution of
the equation
\begin{equation}
\label{eq1} \frac{dx}{dt} = \omega ,{\begin{array}{*{20}c}
 \hfill & \hfill \\
\end{array} }x(0) = \theta,
\end{equation}
which relates to the dynamical system
\begin{equation}
\label{eq2} \Phi (t,c) = ce^{i\omega {\begin{array}{*{20}c}
 t \hfill & \hfill \\
\end{array} }}
\end{equation}
on the circle $\mbox{S}^1 = U(1) = \{e^{i\zeta }\left| {\zeta \in
} \right.\mbox{R}\}$.

In terms of Lie groups, $\mbox{S}^1 = U(1)$ is an abelian Lie group
with its Lie algebra $i\mbox{R} = \{i\omega \left| {\omega \in
\mbox{R}} \right.\},$ and the map
\begin{equation}
\exp :i\mbox{R} \to \mbox{S}^1,{\begin{array}{*{20}c}
 \hfill & \hfill & \hfill \\
\end{array} }\exp (i\omega ) = e^{i\omega },
\end{equation}
is the exponential map of $\mbox{S}^1$. The system (3) is also
called the left-invariant flow, of which the tangent vector field is
called left-invariant vector field on $\mbox{S}^1$.

This fact motivates the idea that the left-invariant flows on the
general Lie groups can be considered as \textit{general
oscillators}. In particular, the left-invariant flows on the
non-Abelian Lie groups can be considered as \textit{non-Abelian
oscillators}. Let $G$ be a Lie group and $\Gamma$ its Lie algebra.
If $v \in \Gamma$, then the left-invariant flow determined by $v$
on the $G$ is
\begin{equation}
\label{eq3} \Phi (t,g) = g \cdot \exp tv,{\begin{array}{*{20}c}
 \hfill & \hfill \\
\end{array} }t \in \mbox{R},{\begin{array}{*{20}c}
 \hfill & \hfill \\
\end{array} }g \in G.
\end{equation}
Different from those defined on the linear Euclidean spaces, the
above dynamical system (5) is defined on the manifold $G$.

Next we consider the synchronization of network, of which each
node is located a general oscillator on the Lie group $G$ with its
Lie algebra $\Gamma$. The collective synchronization behavior of
the coupled general oscillators starts by randomly choosing an
element $v^\alpha \ne 0$ in $\Gamma$ and an initial phase $g_0 \in
G$ for each node $\alpha$. In the case of the general oscillator
$v^\alpha$ and $g_0$ correspond to the frequency $\omega$ and $c =
e^{i\theta}$ in Eq. (3), respectively.

It is worthwhile to note that, in general, the phase space of our
general oscillator is both non-Abelian group in algebra, and
non-linear space, which is a manifold, in geometry. Thus one may
expect that there will be more interesting phenomena in this model.

\begin{figure}
\scalebox{0.8}[0.8]{\includegraphics{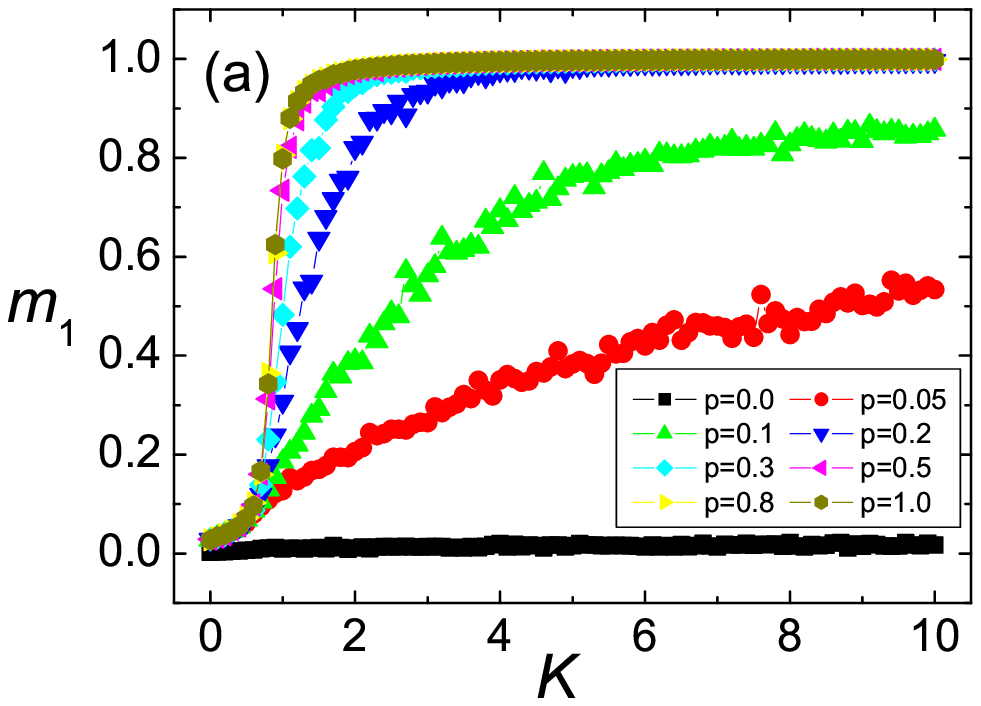}}
\scalebox{0.8}[0.8]{\includegraphics{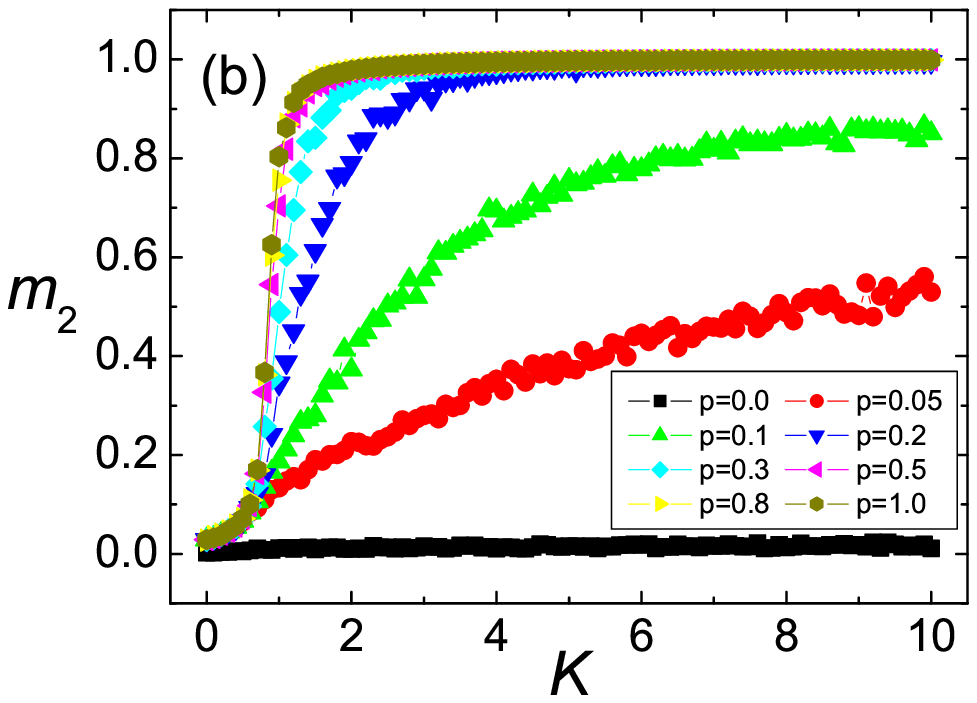}} \caption{(Color
online) Order parameters $m_1$ (a) and $m_2$ (b) vs coupling
strength $K$ for different values of the rewiring probability $p$.
All the simulation results are the average over 100 independent
runs corresponding to the case of $n=2$.}
\end{figure}

For a dynamical system on the manifold $G$, its equations of
motion should be written in a local coordinate system because,
generally, there does not exist a global coordinate system on a
manifold. Let $(x_1 ,x_2 , \cdots, x_n )$ be a local coordinate
system about the identity $e$ of Lie group $G$, $n = \dim G$, and
$(0, \cdots, 0)$ the coordinates of $e$. If $(x_1,x_2, \cdots, x_n
)$, $(y_1 ,y_2 , \cdots, y_n )$ and $(z_1 ,z_2 , \cdots, z_n )$
are the coordinates of $g_1 ,g_2 \in G$ and $g_1 g_2 \in G$ in
this coordinate system respectively, then the multiplication
function $f = (f_1 ,f_2 , \cdots, f_n )$ can be defined as:
\begin{equation}
\begin{array}{l}
z_1 = f_1 (x_1 , \cdots x_n ;y_1 , \cdots y_n )\\
\cdots\\
z_n = f_n (x_1 , \cdots x_n ;y_1 , \cdots y_n ).
\end{array}
\end{equation}
Suppose that $v \in \Gamma$, then the equations of the
left-invariant flow determined by $v$ are
\begin{equation}
\frac{dx_j (t)}{dt} = \sum\limits_i {a_i l_i^j (x(t))}, \quad 1
\le j \le n,
\end{equation}
where $(a_1 ,a_2 , \cdots a_n )$ is the coordinates of $v$ with
respect to the basis $\{\frac{\partial }{\partial x_1
},\frac{\partial }{\partial x_2 }, \cdots ,\frac{\partial
}{\partial x_n }\}_{x = 0} $, and $l = (l_i^j )$ is the matrix
valued function:
\begin{equation}
l_i^j (x) = \frac{\partial f_j (x,y)}{\partial y_i }\vert _{y = 0}
, \quad 1 \le i,j \le n.
\end{equation}
The set of equations (7) is the intrinsic dynamic of a single
general oscillator. Note that, the multiplication functions are
the local representation of the multiplication in the Lie group,
and the matrix valued function (8) is the map induced by
multiplying by $x$ at left. This map translates every vector in
the Lie algebra to $x$. Eqs. (7) mean that the left-invariant flow
determined by $v$ are made from the vector field whose value at
$x$ is just the vector into which translating $v$.

Similar to the case of coupled simple oscillators \cite{Hong2002},
the coupled dynamics of general oscillators are:
\begin{equation}
\begin{array}{l}
 \frac{dx_j^\alpha (t)}{dt} = \sum\limits_i {a_i^\alpha l_i^j (x^\alpha
(t))} - \frac{1}{k_\alpha }\sum\limits_{\beta \in \Lambda _\alpha
}{q_j \left( {x^\alpha \left( t \right) - x^\beta \left( t \right)} \right)} \\
 x_j^\alpha (0) = x_{j0}^\alpha,
 \end{array}
\end{equation}
where $q_j$ is the coupling function assumed smooth, $\Lambda
_\alpha$ is the set of $\alpha$'s neighboring nodes, and
$x_{j0}^\alpha$ is the coordinates of initial phase of node
$\alpha$. According to the existence and uniqueness theorem of
ordinary differential equations, Eq. (9) has an unique solution.
Then, by the theory of Lie groups, this solution can be extended to
an one-parameter subgroup, which corresponds to a left-invariant
flow.

\section{Simulation results for some typical examples}
In this section, we will show some analyses and simulations on
several typical examples. All the simulations are obtained based
on the Watts-Strogatz (WS) network \cite{Watts1998}, which can be
constructed by starting with one-dimensional lattice and randomly
moving one endpoint of each edge with probability $p$. The network
size $N=1000$ and average degree $\langle k \rangle=6$ are fixed.
In our numerical simulation below, the coordinates $a_i$ are
chosen randomly and uniformly in the range $(-0.5, 0.5)$, and the
initial phases $x_{j0}^\alpha$ in $(0, 2\pi)$. All the numerical
results are obtained by integrating the dynamical equations using
the Runge-Kutta method with step size 0.01. The order parameters
(see below) are averaged over 2000 time steps, excluding the
former 2000 time steps, to allow for relaxation to a steady state.

\subsection{$T^n$, the $n$-dimensional torus}
We firstly investigate the $n$-dimensional torus, $T^n =
\underbrace {\mbox{S}^1\times \mbox{S}^1\times \cdots \times
\mbox{S}^1}_n$, which is a connected compact Lie group. Denote
$G_1=T^n$ and its Lie algebra $\Gamma_1 = \underbrace
{i\mbox{R}\times i\mbox{R}\times \cdots \times i\mbox{R}}_n$ with
$[,] \equiv 0$ (i.e. Abelian). This is a direct extension of the
simple Kuramoto oscillators, and will degenerate to Karumoto
oscillators when $n=1$. It follows that an element in $\Gamma_1 $,
that is an intrinsic frequency, is $v = (i\omega _1 ,i\omega _2 ,
\cdots ,i\omega _n )$.

\begin{figure}
\scalebox{0.8}[0.8]{\includegraphics{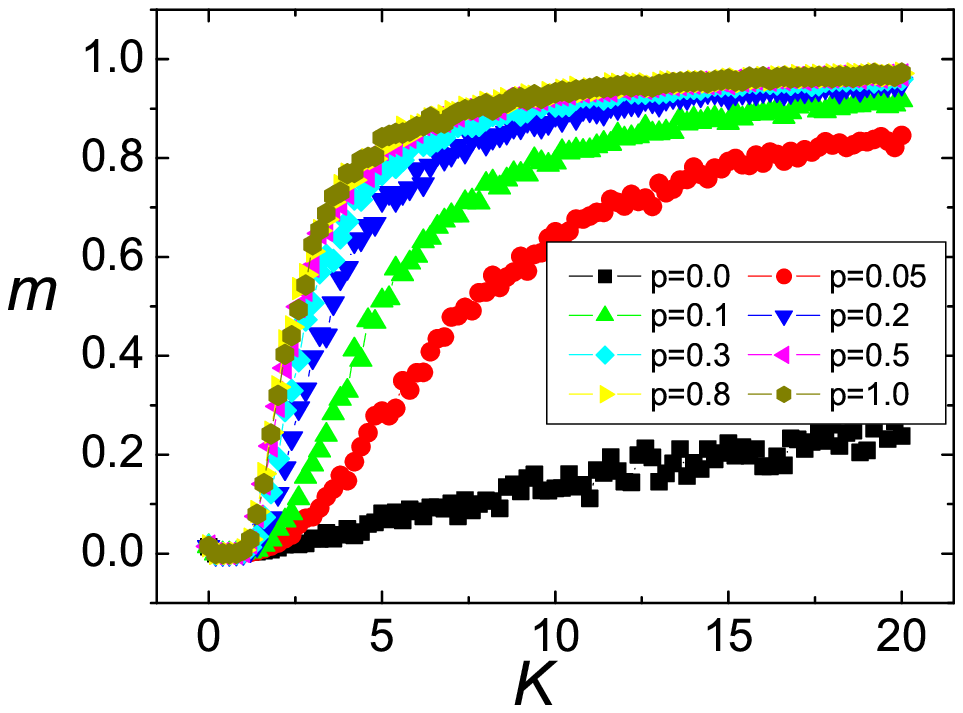}} \caption{(Color
online) Order parameter $m$ vs the coupling strength $K$. All the
simulation results are obtained by averaging 100 independent runs.}
\end{figure}

It is easy to see that the general oscillation determined by $v$
is periodic if $\omega _1 ,\omega _2 , \cdots \omega _n $ are
integer-linearly dependent, and quasi-periodic if $\omega _1
,\omega _2 , \cdots \omega _n $ are integer-linearly independent.
For the equations of motion we have
\begin{equation}
\label{eq6} \frac{d}{dt}\left( {\begin{array}{l}
 x_1^\alpha \\
 \vdots \\
 x_n^\alpha \\
 \end{array}} \right) = \left( {\begin{array}{l}
 \omega _1^\alpha \\
 \vdots \\
 \omega _n^\alpha \\
 \end{array}} \right) - \frac{K}{k_\alpha }\sum\limits_{\beta \in \Lambda
_\alpha } {\left( {\begin{array}{l}
 \sin (x_1^\alpha - x_1^\beta ) \\
 \vdots \\
 \sin (x_n^\alpha - x_n^\beta ), \\
 \end{array}} \right)}
\end{equation}
with the initial conditions $x_j^\alpha \left( 0 \right) =
x_{j0}^\alpha$, where $K$ is the coupling strength. To
characterize the synchronized states of the $j$th dimension, we
use the order parameter
\begin{equation}
m_j=\left\{\left|\frac{1}{N}\sum_\alpha
e^{ix_j^\alpha}\right|\right\},\texttt{ } j=1,2,\cdots,n,
\end{equation}
where $\{\cdot\}$ signifies the time averaging.

Fig. 1 displays the relations between the phase order parameter and
the coupling strength for various values of the rewiring probability
$p$, where $n=2$. Since the dynamical behaviors in different
dimensions are independent, and each one is equal to a simply
Kuramoto oscillator, the synchronization behavior in each dimension
is the same as that of a Kuramoto oscillator \cite{Hong2002}. In a
word, the small-world effect ensures the global synchronization for
sufficiently large coupling strength, and the higher disorder (i.e.
larger $p$) will enhance the network synchronizability.

This example can be used to describe the collective behavior of
the coupled double planar pendulums in a network. Furthermore,
according to the Liouville-Arnold theorem, the phase space of a
integrable system, if it is compact, is a $n$-torus. Therefore,
this example can be widely applied to describe the synchronization
phenomenon of integrable systems.

\subsection{$GL_0(3,R)$, the identity component of 3-dimensional general linear group $GL(3,R)$}
$GL_0(3,R)$, the identity component of $GL(3,R)$, is a connected
Lie group of dimension 9 (Actually, $GL_0(3,R)$ is a open subset
of $R^9$). Denote $G_2=GL_0(3,R)$, and $\Gamma_2$ its Lie algebra
containing all the $3\times 3$-real matrices. The coordinates
$(g_{ij})$ for all $g = (g_{ij} ) \in G_2$ can be addressed on the
basis $E_{ij}$ of $\Gamma_2 $ (see the case (a) of Appendix A),
where
\[
E_{11} = \left( {{\begin{array}{*{20}c}
 1 \hfill & 0 \hfill & 0 \hfill \\
 0 \hfill & 0 \hfill & 0 \hfill \\
 0 \hfill & 0 \hfill & 0 \hfill \\
\end{array} }} \right),
\quad
 \cdots ,
\quad E_{33} = \left( {{\begin{array}{*{20}c}
 0 \hfill & 0 \hfill & 0 \hfill \\
 0 \hfill & 0 \hfill & 0 \hfill \\
 0 \hfill & 0 \hfill & 1 \hfill \\
\end{array} }} \right).
\]

In this circumstances the equations of motion are
\begin{equation}
\frac{dx_{ij}^\alpha (t)}{dt} = \sum\limits_{p = 1}^3
{x_{ip}^\alpha (t)a_{pj}^\alpha } - \frac{K}{k_\alpha }\sum_{\beta
\in \Lambda_\alpha} {(x_{ij}^\alpha (t) - x_{ij}^\beta (t))},
\end{equation}
with the initial conditions $x_{ij}^\alpha (0) = x_{ij0}^\alpha$,
where $(a_{ij}^\alpha ) \in \Gamma_2 $ and $K$ is the coupling
strength.

\begin{figure}
\scalebox{0.8}[0.8]{\includegraphics{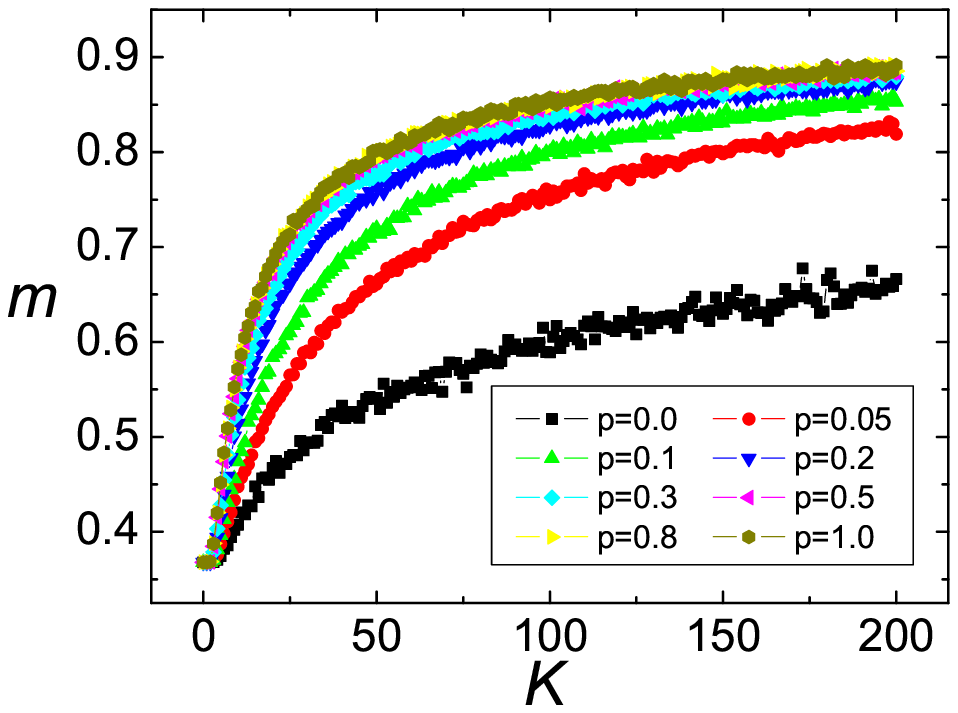}} \caption{(Color
online) Order parameter $m$ vs the coupling strength $K$. All the
simulation results are obtained by averaging 100 independent runs.}
\end{figure}

Since $G_2$ is a subset of the Euclidean space $R^9$, the coupling
term of arbitrary pair of nodes $\alpha$ and $\beta$ can be directly
written as $x_{ij}^\alpha-x_{ij}^\beta$. Therefore, the order
parameter is defined as:
\begin{equation}
\label{eq8} m = \left[ {\left\langle {\frac{2}{N(N -
1)}\sum\limits_{\alpha < \beta } {\exp \left[ { - c\left\| {\tilde
{x}^\alpha - \tilde {x}^\beta } \right\|^2} \right]} }
\right\rangle } \right],
\end{equation}
where $\tilde {x}^\alpha = (\tilde {x}_{ij}^\alpha ) =
\frac{\left( {x_{ij}^\alpha } \right)}{M}$ and $M = \max\left\{
{\left\| {x^\beta (t)} \right\|} \right\}_{1 \le \beta \le N}$.

In Fig. 2, we report the simulation results about the order
parameter $m$ as a function of $K$ for different rewiring
probabilities. Similar to the situation of $T^n$, this system can
approach to the global synchronized state for sufficiently large
$K$ when $p$ is large enough. It is worthwhile to emphasize that
the order parameter of different systems can not compare to each
other directly, since their definitions are not the same.

\subsection{$SU(2)$, the special unitary group}
$SU(2) = \mbox{S}^3$, the special unitary group, is a connected
compact Lie group consisted of all anti-Hermitian traceless
matrices. Denote $G_3=SU(2)$, and $\Gamma_3$ its Lie algebra having
the following basis
\[
J_1 = \frac{1}{2}\left( {{\begin{array}{*{20}c}
 i \hfill & 0 \hfill \\
 0 \hfill & { - i} \hfill \\
\end{array} }} \right),
\quad J_2 = \frac{1}{2}\left( {{\begin{array}{*{20}c}
 0 \hfill & 1 \hfill \\
 { - 1} \hfill & 0 \hfill \\
\end{array} }} \right),
\quad J_3 = \frac{1}{2}\left( {{\begin{array}{*{20}c}
 0 \hfill & i \hfill \\
 i \hfill & 0 \hfill \\
\end{array} }} \right).
\]
Give an arbitrary element $g \in SU(2)$, its local coordinates $(x_1
,x_2 ,x_3 )$ can be found from:
\begin{equation}
\label{eq9} g = \left( {{\begin{array}{*{20}c}
 {w_1 } \hfill & {w_2 } \hfill \\
 { - \bar {w}_2 } \hfill & {\bar {w}_1 } \hfill \\
\end{array} }} \right) = (\exp x_1 J_1 )(\exp x_2 J_2 )\left( {\exp x_3 J_3
} \right),
\end{equation}
and
\begin{equation}
\label{eq10} w_1 \bar {w}_1 + w_2 \bar {w}_2 = 1.
\end{equation}

The corresponding equations of motion are
\begin{equation}
\frac{dx_j^\alpha (t)}{dt} = \sum\limits_{i = 1}^3 {a_i^\alpha l_i^j
(x^\alpha (t))} - \frac{K}{k_\alpha }\sum\limits_{\beta \in
\Lambda_\alpha} {\sin \frac{1}{2}(x_j^\alpha (t) - x_j^\beta (t))},
\end{equation}
with the initial condition $x_j^\alpha (0) =
x_{j0}^\alpha,\texttt{}j=1,2,3$, where $K$ is the coupling strength.
It is worthwhile to emphasize that the behaviors of this dynamical
system is independent to the local coordinate systems. It follows
that the local coordinates (14) in $SU(2)$ can be replaced by
\begin{equation}
g = \exp \left( {x_1 J_1 + x_2 J_2 + x_3 J_3 } \right),
\end{equation}
which is convenient for the description of two-state quantum systems
since the matrices $J_1 $,$J_2 $, and $J_3 $ are relative to Pauli
matrices as:
\begin{equation}
-2iJ_1=\sigma_3,\quad -2iJ_2=\sigma_2,\quad -2iJ_3=\sigma_1.
\end{equation}

Note that, a one-parameter subgroup $U(t)\subset SU(2)$ can be
considered as the time-evolution in the quantum mechanics. Since the
subgroup $e^{i\zeta}U(t)$, where $\zeta$ is an arbitrary real
number, represents the same time-evolution as $U(t)$ does, we call
two different nodes $\alpha$ and $\beta$ are synchronous if they are
only differ in a phase factor $e^{i\zeta}$, that is
\begin{equation}
\left( {{\begin{array}{*{20}c}
 {w_1^\alpha } \hfill & {w_2^\alpha } \hfill \\
 { - \bar {w}_2^\alpha } \hfill & {\bar {w}_1^\alpha } \hfill \\
\end{array} }} \right)\left( {{\begin{array}{*{20}c}
 {w_1^\beta } \hfill & {w_2^\beta } \hfill \\
 { - \bar {w}_2^\beta } \hfill & {\bar {w}_1^\beta } \hfill \\
\end{array} }} \right)^{ - 1} = \left( {{\begin{array}{*{20}c}
 {e^{i\zeta }} \hfill & 0 \hfill \\
 0 \hfill & {e^{ - i\zeta }} \hfill \\
\end{array} }} \right),
\quad \zeta \in \mbox{R}.
\end{equation}
Therefore, if the equation
\begin{equation}
w_1^\alpha w_2^\beta = w_2^\alpha w_1^\beta
\end{equation}
is hold for every $1 \le \alpha < \beta \le N$, the completely
global synchronization is achieved. Consequently, for each pair of
nodes $\alpha$ and $\beta$, define the dynamical ``distance" between
them as
\begin{equation}
\mu _{\alpha \beta } = \left| {w_1^\alpha w_2^\beta - w_2^\alpha
w_1^\beta } \right|,
\end{equation}
where $|\cdot|$ signifies the modulus of the complex number.
Accordingly, the order parameter is
\begin{equation}
m_3 = \left[ {\left\langle {\exp \left( { - \max \left\{ {\mu
_{\alpha \beta } \left| {1 \le \alpha < \beta \le N} \right.}
\right\}} \right)} \right\rangle } \right].
\end{equation}

In Fig. 3, we report the simulation results about the order
parameter $m$ as a function of $K$ for different rewiring
probabilities. Analogously, this system can approach to the global
synchronized state for sufficiently large $K$ when $p$ is large
enough. Note that, $e^{ - 1} \le m \le 1$ for $0 \le \left| {\mu
_{\alpha \beta } } \right| \le 1$. Thus in the uncoupling case (i.e.
$K=0$), the order parameter $m$ approaches to 0.37.

In terms of the Hopf bundle $\mbox{S}^1 \to \mbox{S}^3\buildrel p
\over \longrightarrow \mbox{S}^2$, the base space $\mbox{S}^2$ is
the set of states of a two-state quantum system (see the ref.
\cite{Frankel2004} and Appendix B for details). Therefore, let a
two-state quantum system be located at each node of a small-world
network, the whole system can achieve the synchronized state if the
coupling strength is large enough. Note that, the special orthogonal
group $SO(3)$ has an isomorphic Lie algebra to $\Gamma_3$ for the
existence of the double covering $SU(2) \to SO(3)$, which is a
homomorphism. Therefore, the dynamical system $SO(3)$ has the same
synchronized behavior as that of $SU(2)$. The case where each node
is located with a top can be represented by the system $SO(3)$
\cite{Arnold1978}.

\section{Conclusion}
In this paper, we extended the concept of Kuramoto oscillator to the
left-invariant flow on general Lie group, and studied the
generalized phase synchronization on small-world networks. The
left-invariant flow on Lie group can be used to represent many
significant physical systems, such as the integrable systems (e.g.
the double planar pendulums), the two-state quantum systems (e.g.
Ising model), the motions of top, and so on. In particular, the
dynamics of two-state quantum systems on complex networks are
extensively studied recently \cite{Herrero2002,Hong2002,Zhu2004},
the present work can provide us a theoretical frame to investigate
their collective synchronized behaviors.

It is intuitive that the low-dimensional systems are more easily to
synchronize. However, one may note that in the regular networks
(i.e. $p=0$), the system $T^2$ does not exhibit synchronized
behavior while the obvious synchronization is observed in the higher
dimensional systems $GL_0(3,R)$ and $SU(2)$ when the coupling
strength gets large. This is an interesting phenomenon that worths a
further study in the future.

\begin{acknowledgments}
We thank Dr. Hui-Jie Yang for helpful discussions. This work was
partially supported by the National Natural Science Foundation of
China under Grant Nos. 70471084, 70471033 and 70571074, the Special
Research Founds for Theoretical Physics Frontier Problems under
Grant No. A0524701, the Specialized Program under the Presidential
Funds of the Chinese Academy of Science, and the Graduate Student
Foundation of USTC under Grant No. KD2005007.
\end{acknowledgments}

\appendix

\section{How to choose the coordinate system}
In order to investigate the collective synchronization behaviors of
the coupled non-Abelian oscillators, we have to integrate Eq. (9)
numerically. Because of the using of local coordinate system, a
question rises: Does the Eq. (9) express the behavior of the system
in the whole manifold $G$? The answer is as follows.

\underline {Case (a)} Manifold $G$ is a connected open subset of
Euclidean space $R^n$. In this case, we can directly use the natural
coordinate system of $R^n$ as the local coordinate system of $G$.

\underline {Case (b)} $G$ is a connected compact Lie group. In this
case, we have $\exp (\Gamma) = G$, so we can choose the coordinate
system of $\Gamma$ as the local coordinate system of $G$.

\section{Relevance of the dynamical system $SU(2)$}
According to the theory of fiber bundles, the condition (20) is
equivalent to that $g^\alpha(t)$ and $g^\beta(t)$ is all in the same
fiber of the Hopf bundle $\mbox{S}^1 \to \mbox{S}^3\buildrel p \over
\longrightarrow \mbox{S}^2$. In order to interpret the map $p$ we
consider the 2-sphere $\mbox{S}^2$ as the complex projective line
$\mbox{CP}^1$ in which a point is written as $[w_1,w_2]$, where
$w_1$ and $w_2$ are two complex numbers and not all zero. Here, the
point $[w_1,w_2]$ is identified with $[\lambda w_1,\lambda w_2]$ for
any nonzero complex number $\lambda$. Then the map $p$ can be
written as
\begin{equation}
p(w_1,w_2)=[w_1,w_2].
\end{equation}
It is easy to show that the inverse image $p^{-1}[w_1,w_2]$ is the
sphere $\mbox{S}^1$ \cite{Frankel2004}. This fact tells us that, in
the sense of our definition, the collective synchronization
behaviors of the coupled non-Abelian oscillators $SU\left( 2
\right)$ can represent the two-state quantum system.


\begin{thebibliography}{Strogatz2003}
\bibitem{Strogatz2003} S. H. Strogatz, {\it SYNC-How the emerges from chaos in the universe, nature, and daily life} (Hyperion, New York, 2003).
\bibitem{Kuramoto1984} Y. Kuramoto, \emph{Chemical Oscillations, Wave and Turbulence} (Springer-Verlag, Berlin, 1984).
\bibitem{Pikovsky2001} A. Pikovsky, \emph{Synchronization} (Cambridge University Press, Cambridge, 2001).
\bibitem{Acebron2005} J. A. Acebr\'on, L. L. Bonilla, C. J. P. Vicente, F. Ritort, and R. Spigler, Rev. Mod. Phys. {\bf 77}, 137 (2005).
\bibitem{Schutz1980} B. F. Schutz, {\it Geometrical method of mathematical physics} (Cambridge University Press, 1980).
\bibitem{Brocker1985} T. Brocker, and T. T. Dieck, {\it Representations of compact Lie groups} (Springer-Verlag, 1985).
\bibitem{Hong2002} H. Hong, M. Y. Choi, and B. J. Kim, Phys. Rev. E {\bf 65}, 026139 (2002).
\bibitem{Watts1998} D. J. Watts, and S. H. Strogatz, Nature {\bf 393}, 440 (1998).
\bibitem{Frankel2004} T. Frankel, {\it The Geometry of Physics, An Introduction} (Cambridge University Press, Cambridge, 2004).
\bibitem{Arnold1978} V. I. Arnold, {\it Mathematical Methods of Classical Mechanics} (Springer-Verlag, New York, 1978).
\bibitem{Herrero2002} C. P. Herrero, Phys. Rev. E {\bf 65}, 066110 (2002).
\bibitem{Hong2002} H. Hong, B. J. Kim, and M. Y. Choi, Phys. Rev. E {\bf 66}, 018101 (2002).
\bibitem{Zhu2004} C. -P. Zhu, S. -J. Xiong, Y. -J. Tian, N. Li, and K. -S. Jiang, Phys. Rev. Lett. {\bf 92}, 218702 (2004).
\end{thebibliography}
\end{document}